\documentclass[12pt,a4paper]{article}

\usepackage[left]{lineno}
\usepackage{color}
\usepackage[utf8]{inputenc}
\usepackage{amsmath}
\usepackage{graphicx}
\usepackage{arydshln}
\usepackage{setspace}
\usepackage{overcite}
\usepackage{titling}
\usepackage{bm}

\usepackage{caption}
\usepackage{authblk}
\usepackage{siunitx}
\usepackage[top=30truemm,bottom=30truemm,left=25truemm,right=25truemm]{geometry}
\usepackage{url}

\DeclareSIUnit\Molar{M}

\pretitle{\begin{flushleft}\Large \textbf}
\posttitle{\end{flushleft}}
\preauthor{\begin{flushleft}}
\postauthor{\end{flushleft}}

\title{\textit{C. elegans} collectively forms dynamical networks}
\author[1]{\large Takuma Sugi}
\author[2]{Hiroshi Ito}
\author[1]{Masaki Nishimura}
\author[3]{Ken H. Nagai}
\author{}

\affil[1]{Molecular Neuroscience Research Center, Shiga University of Medical Science, Ohtsu, Shiga 520-2121, Japan}
\affil[2]{Faculty of Design, Kyushu University, Fukuoka 815-8540, Japan}
\affil[3]{School of Materials Science, Japan Advanced Institute of Science and Technology, Nomi, Ishikawa 923-1292, Japan}
\date{}

\begin{document}
\maketitle

{\bf Ordered collective motions are ubiquitous among locally interacting living beings~\cite{Couzin_collective_2009,Kondo:2010bx,Vicsek:2012gp} and play significant roles in biological functions such as biofilm formation,~\cite{Zhao2013} wound healing,~\cite{Poujade2007} the flocking of birds and the schooling of fish.~\cite{Hemelrijk2012, Calovi2014, katz_inferring_2011} Research on the physics of active matter, which seeks to identify unified descriptions of such collective motions, has included various studies using mathematical models with only simple rules of self-propulsion~\cite{Reynolds1987,Vicsek:1995ti,Vicsek:2012gp}. Experimental systems have been developed for non-living self-propelled particles (SPPs)~\cite{Sumino:2012dw, Suzuki:2015fs, Bricard2013}, bacteria,~\cite{Zhao2013, Nishiguchi2017} and mammalian cells on a substrate,~\cite{Kawaguchi:2017em,saw_topological_2017} but there have been no experimental systems of multicellular organisms, which have much more complex behaviours~\cite{Hebb:1949vd, Kandel:2001is}, with various controllable parameters over a wide range.~\cite{Popkin:2016cy} Thus, whether the collective motions of animals can be described by minimal models remains unknown. Here, we have constructed an experimental system in which the conventional model animal \textit {C. elegans} collectively forms dynamical networks of bundle-shaped aggregates. We investigated the dependence of our designed experimental system on the extrinsic parameters (the substrate, ambient humidity and density of worms). Taking advantage of \textit {C. elegans} genetics, we also controlled the intrinsic parameters (genetically determined motility) by mutations or by forced neural activation via optogenetics. Furthermore, we developed an agent-based model with simple rules that reproduced the dynamical network formation and its dependences on the parameters, suggesting that the key behaviours for the network formation were alignment of worms after collision and smooth turning. This result revealed that the collective motion of living things with advanced information processing systems can be described by a simple minimal model in a broad area of parameter space. Our results will pave the way for the understanding of biological functions of groups of animals via the concepts of active matter physics.}
 
 Some species of nematodes, particularly parasitic nematodes, have long been known to aggregate and swarm on their habitats to survive desiccation for extended periods\cite{GRAY:1964ti,Gaugler:2004uj}. This swarming was quantitatively proven to be one of the most important survival strategies for adaptation to the demands of fluctuating environments that occasionally become extreme and life-threatening~\cite{Crowe:1975gu,Higa:1993gg,Erkut:2015cn}. The free-living nematode \textit {C. elegans}, a commonly used laboratory model animal, is genetically tractable and thus potentially offers a great opportunity to experimentally investigate the collective motions of animals under the control of a wide variety of parameters. However, \textit {C. elegans} researchers have paid little attention to establishing an experimental system for producing large-scale \textit {C. elegans} swarms~\cite{Gaugler:2004uj} because it is difficult to prepare very large numbers of worms on a solid surface in the standard cultivation protocol of \textit {C. elegans} due to the decrease in its propagation speed as food becomes limited. To overcome this difficulty, we applied a culture method previously established for other nematodes, in which dog food agar (DFA) medium was used~\cite{Tanaka:2010hz,united1981monoxenic}. DFA medium contains enough nutrients to enable the propagation of a very large number of worms for a long time ($>$ one month). During propagation, the worms were kept in the dauer state, a starvation-like state induced by crowded conditions~\cite{Tanaka:2010hz,united1981monoxenic,Cassada:1975tf}. In a week of cultivation in DFA medium in a glass bottle, the propagated \textit {C. elegans} sometimes climbed up the inner wall of the bottle. We noticed that when the bottom of a bottle was warmed from 23ºC to 25ºC, large numbers of propagated worms climbed up the inner surface of the bottle. We then observed the emergence of a dynamical network structure on the inner surface of the glass bottles (Fig. 1a, b and Supplementary Video 1). The structure comprised a large number of compartments surrounded by bundle-shaped worm aggregates. The average diameter of the compartments was clearly larger than the typical body size of a single worm (431 $\pm$ 64 \si{\micro\meter} (s.d., n = 13)). The bundle in the magnified image in Fig. 1b was highly crowded with worms, indicating that the effect of excluded volume was significant. The network pattern dynamically remodelled over time by the repeated coalescence and division of compartments in approximately 100 \si{\second} (Fig. 1c-f). 
 
We found that the worms could form a qualitatively identical network on a plastic substrate by climbing up to the inner surface of the lid from DFA on agar medium inside a Petri plate (Fig. 1g, Extended Data Fig. 1). We also examined whether the network could be formed on a food (\textit{Escherichia coli} OP50)-free agar surface of nematode growth medium (NGM), which was used in the standard cultivation method instead of the naturally habitable soil environment~\cite{Brenner:1974wn}. For this experiment, we collected large numbers of worms that were propagated in DFA medium and climbed up to the inner surface of the lid using water, and then the obtained droplet was placed on an agar surface. We observed that the worms also formed a dynamical network after water seeped into the agar medium (Fig. 1h). The qualitatively identical collective motions on three different substrates indicate that \textit {C. elegans} forms this dynamical network regardless of the substrate material as long as the number of worms is large enough. The network formation is noted to require self-propelled activity, as no obvious self-organized pattern was observed for dead worms on agar (Fig. 1i). 
 
This reconstructed system of living animals enables us to easily control certain extrinsic parameters, such as the density of worms and the humidity in a Petri plate. To control the density, we collected propagated worms from the inner surfaces of lids and prepared several 200 \si{\micro\litre} droplets with worm densities in the range of 500 to 2,000 worms/\si{\micro\litre}. Each of the droplets was placed on the agar surface. We found that the formation of the dynamical network required a worm density greater than 1,000 worms/\si{\micro\litre} (200,000 worms on the agar surface) (Fig. 2a). We also controlled the humidity as another extrinsic parameter. To increase the humidity around the worms, a plate with agar in the bottom and worms on the lid was heated from 23ºC to 26ºC from the bottom at a speed of ca. 1ºC per min (Extended Data Fig. 1). The heating evaporated the water in the agar to the air inside the Petri plate, thereby gradually increasing the humidity. As previously reported for another nematode, \textit {Panagrellus redivivus}~\cite{Gart:2011ez}, we confirmed that the attachment of colliding worms was driven by the surface tension of the water around the worms, and this attractive force was strong enough to keep the worms in an aggregated state in opposition to the repulsive force exerted by their muscles. The increase in humidity strengthened the attraction (Supplementary Video 2). The time evolution of the network on the lid is shown in Fig. 2b and Supplementary Video 3. A clear dynamical network was observed at the beginning of the experiment (the upper row of Fig. 2b (0 min)). Then, as the humidity increased, the compartments grew larger (the upper row of Fig. 2b (11.4 min and 19.1 min) and Supplementary Video 3). Further increasing the humidity led to the collapse of the network structure and the formation of many simple aggregates (the upper row of Fig. 2b (23.2 min)). For a while, the shape of the aggregates largely fluctuated and small running aggregates sometimes split from protuberances at the edges of the aggregates (Supplementary Video 4). Finally, all the aggregates became still. 

 The accumulated genetic resources and tools, especially related to motility, that have been developed in \textit {C. elegans}~\cite{Corsi:2015bv} enabled easy control of the intrinsic parameters of the collective motion of the worms. We examined the collective motion of the \textit{mec-4(e1611)} mutant, which carries a mutation in the gene encoding the mechanosensory Na$^+$ channel MEC-4 and moves along a circular trajectory with higher curvature than that of the wild-type worm~\cite{Cohen:2012ec}. We cultivated the \textit{mec-4(e1611)} mutant in DFA on NGM inside a Petri plate and increased the humidity inside the plate, as we did for the wild-type worms (the lower row of Fig. 2b and Supplementary Video 5). Subsequently, we observed similar network formation by the \textit{mec-4(e1611)} mutants during changes in humidity, but its compartment size was smaller than that of the wild-type worms. In addition, we constructed an experimental system for optical manipulation of the network. This newly established optogenetic system (see Methods) enables a transient perturbation to the worm aggregations by driving the movement of halted worms (approximately 20\% of worms before light illumination) by the optical activation of mechanosensory neurons. As soon as the light was turned on, the bundle started to collapse, which suggested that forcing halted worms to move destabilized the bundle. We found that the bundle could recover to the original shape as long as the duration of activation was not long ($<$ 2 \si{\second}) (Supplementary Video 6). Activation with mild light intensity for a prolonged duration ($>$ 30 \si{\second}) broke the bundle. After the light was turned off, the worms formed different bundles (Supplementary Video 7). It is noted that without all-trans-retinal (ATR), which is needed for the optogenetic activation of the target neuron, no response of the bundles to light illumination was observed (Supplementary Video 8).
 
To elucidate the key behavioural characteristics necessary for network formation, behavioural data were collected at the single-worm level. We analysed the pair interactions and trajectories of sparsely isolated worms ($\sim$ 1 worm/\si{\square\milli\metre}) on the inner plastic surface of the lid of an NGM plate. A large number of random collisions occurred between worms in close proximity. In the 43 pair collision events, the outgoing angles were near $0$ or $\pi$, irrespective of the value of the incoming angles (Fig. 3a). This result indicates that collisions induce the alignment or anti-alignment of worms, i.e., nematic rather than polar order. We found that individual \textit {C. elegans} in the network were also nematically aligned by using transparent worms labelled with a fluorescent protein TagRFP. A total of 51.3\% of worms moved in the directions $0 \pm \pi/12$ rad (28.2\%) and $\pi \pm \pi/12$ rad (23.1\%) along a bundle (Fig. 3b and 3c). The isolated dauer worms exhibited clockwise or anti-clockwise circular trajectories with a gradual change in rotation rate, as observed for fed worms in a previous study~\cite{Cohen:2012ec} (Fig. 3d and Supplementary Video 9). These trajectories included short-wavelength oscillation, which arises from the worms’ undulatory locomotion. From the analysis of 38 worms (see Methods section), we obtained the mean velocity ($v_0$ = 87 $\pm$ 30 (s. d.) \si[per-mode=symbol]{\micro\meter\per\second}) and the mean rotation rate ($\omega_0 = 1.4 \times 10^{-3} \pm 1.9 \times 10^{-2}$ (s.e.) rad/s). The standard deviation and the correlation time of the rotation rate were estimated to be $\sigma_{\omega} = 0.155$ (95 percent confidence interval, 0.151 to 0.159) rad/s and $\tau = 27$ (95 percent confidence interval, 25 to 29) s, respectively. 

The short-range nematic alignment and smooth turning of \textit {C. elegans} are reminiscent of microtubules driven by axonemal dynein c. The hexagonal lattice of vortices formed by the microtubules was reproduced by a simple agent-based model, in which the agents have a memory of the rotation rate\cite{Sumino:2012dw}. To confirm whether the collective motion of \textit {C. elegans} could be reproduced by the minimal model, we employed one based on the model in Refs~\cite{Sumino:2012dw} and \cite{nagai_collective_2015}. In addition to short-range nematic interaction and smooth turning, two additional significant characteristics, the attraction caused by surface tension ($\bm{F}^{\rm a}$) and the repulsion due to the excluded volume of the nematodes ($\bm{F}^{\rm r}$), were considered. The simulated equations were
\begin{eqnarray*}
\dot{\bm{r}_{i}}&=&\bm{e}_{\theta_{i}}+\sum_{{r_{ij}<r^{\rm r}}}{\bm{F}^{\rm r}_{ij}}+\sum_{{r^{\rm r}<r_{ij}<1}}{\bm{F}^{\rm a}_{ij}}\\
\dot{\theta}_i&=&\omega_i+\frac{1}{N_i}\sum_{r^{\rm r}<r_{ij}<1}\sin{2(\theta_j-\theta_i})\\
\dot{\omega}_i&=&-\frac{\omega_i-\omega_0}{\tau}+\sqrt{\frac{2}{\tau}}\sigma_{\omega} \xi_i\\
\bm{F}^{\rm r}_{ij}&=& k^{\rm r}(r_{ij}-r^{\rm r})\bm{e}_{ij}\\
\bm{F}^{\rm a}_{ij}&=& \frac{k^{\rm a}}{r_{ij}}\bm{e}_{ij},
\end{eqnarray*}
 where $\bm{r}_{i}$, $\theta_i$, and $\omega_i$ are the position, the direction of motion, and the rotation rate of particle $i$, respectively, and $\bm{e}_{\theta_i}$ is the unit vector in the direction of $\theta_i$, which means that isolated particles move with the speed of 1. The repulsive force, $\bm{F}^{\rm r}_{ij}$, is exerted on particle $i$ by particle $j$ in a circle with a radius of $r^{\rm{r}}$ and its centre at $\bm{r}_i$. The direction of $\bm{F}^{\rm{r}}_{ij}$ was the same as the unit vector from particle $i$ to particle $j$, $\bm{e}_{ij}$. When $r^{\rm{r}} < r_{ij}=\left|\bm{r}_i-\bm{r}_j\right|<1$, an attractive force, $\bm{F}^{\rm a}_{ij}$, is exerted by particle $j$. The inverse of the distance of the particles was chosen as the dependence of $\bm{F}^{\rm a}_{ij}$ on the distance from the neighbours, with reference to Ref~\cite{Kralchevsky2000}. In the area $\bm{F}^{\rm a}_{ij}$ works, particle $i$ aligns head to head or head to tail with particle $j$. The alignment interaction term is normalized by the number of interacting particles to avoid excessively strong interactions. The Euler method with time interval of 0.001 was used to solve the equations numerically. The values of $r^{\rm r}$ and $k^{\rm r}$ were fixed to 0.2, and 10, respectively. 
 
 Since the particle’s speed is 1 and the radius of the interaction range is 1, the parameters corresponding to the experiments were the estimates from single-worm tracking normalized by half the body length and the average time during movement over half the body length. Half the body length was chosen because the worms were shortened due to the undulation of their bodies during crawling. When the average ($\omega_0 = 0.0035$) and standard deviation ($\sigma_{\omega} = 0.35$) of the rotation rate and the rotational correlation time ($\tau = 10$) corresponding to the aforementioned estimated values were used, a dynamical network was successfully formed, as shown in Fig. 4a and Supplementary Video 10. 
 
To compare the dynamical networks of the model and the worms, we measured the area surrounded by thick bundles (see Methods). In the large size range, the histogram of the logarithm of size took the form of a Gaussian distribution in the model case (Fig. 4b). Indeed, the quantile-quantile plot~\cite{WILK1968} of the fitted Gaussian curve vs. the histogram in the range larger than 50 was linear (Inset in Fig. 4b). In the case of the dynamical network of \textit {C. elegans} in Supplementary Video 1, in the size range larger than 0.73 \si{\square\milli\metre}, the histogram of the logarithm of compartment size also took the form of a Gaussian distribution, as shown in Fig. 4c. The same distribution form in the real and the model case indicates that both dynamical networks are governed by a common physical process and that the minimal model reproduced the dynamical network formation of the worms. The log-normal distribution of object size is often observed in random nucleation and growth processes~\cite{Bergmann1998} and in pulverization~\cite{Kolmogorov1941}. When objects are divided into fragments, the division in which the ratio of the original objects and the fragments is distributed independently of the original particle size leads asymptotically to a log-normal size distribution. Thus, the log-normal distribution in the dynamical networks suggests that the compartments divide independently of the area size.
 
By the model, the extrinsic parameter dependence of the network was well reproduced. A sufficiently large density of worms was needed in the model, as in the case of real worms (Figs. 2a and 4d). Fig. 4e shows the $k^{\rm a}$ dependence, which corresponds to the humidity dependence of the network of worms. The upper figures of Fig. 4e was in the case of the model with parameters estimated from single-worm tracking. The size of the compartments gradually increased with increasing $k^{\rm a}$, and non-motile aggregates were formed when $k^{\rm a}$ was larger than the critical value, as in the case with increasing humidity. 

The intrinsic parameter dependence was also compared. As stated above, the \textit{mec-4} mutant formed a dynamical network with smaller compartments (Fig. 2b). In the simulations, the particles travelling with higher $\sigma_{\omega}$ and $\omega_0$ formed the smaller compartments (Fig. 4e). As for the optical stimulation, our model responded to the increase in the number of active particles in a similar manner to the optogenetically induced response. When 30~\% of the particles were inactive, the dynamical network was formed, as in the initial state of Supplementary Video 11. To make the particles inactive, zero was multiplied to the terms $\bm{e}_{\theta_{i}}$ in the first equation and $\omega_{i}$ in the second equation. The sudden change from the behaviour of the inactive particles to that of normal particles collapsed the network structure. Once the ratio of inactive particles returned to its initial value, the network reformed. 

The nematode \textit {C. elegans} was introduced as a genetically tractable laboratory animal~\cite{Brenner:1974wn,Corsi:2015bv}. Since then, behavioural genetic studies using this model animal have contributed to the investigation of individual-level behavioural paradigms. However, over the past 50 years, although a simple clumping pattern was observed~\cite{deBono:1998ic,Artyukhin:2015hd}, no reports have demonstrated dynamical pattern formation via the group-level behaviour of \textit{C.~elegans}. One of the breakthroughs in this study was the use of DFA medium, which allows the easy maintenance of a very large number of worms for an extended time in a Petri plate. Using DFA medium, we have presented the first observation of dynamical collective behaviour by \textit {C. elegans}, thereby introducing a new behavioural paradigm. This finding implies that if a very large number of animals can be maintained, we might find unknown dynamical behaviors much larger than their bodies even in traditionally used model animals, such as larval zebrafish. The great advantage of this experimental system using \textit {C. elegans} is the easy simultaneous observation of the dynamics at both the individual and collective levels thanks to its submillimetre body length. This advantage and the genetic tractability of \textit {C. elegans} offer unprecedented opportunities to control not only extrinsic but also intrinsic parameters, enabling examination of whether the collective behaviours can be described by a simple model. We can state that our results provide a simple physical description of animal collective behaviours. As aggregation is the strategy by which nematodes resist desiccation or forage~\cite{Crowe:1975gu,deBono:1998ic,Artyukhin:2015hd}, we expect that network formation can also be linked to survival strategies. The physiological basis of nematode survival should be reconsidered with the help of the physics of active matter. 

\section*{METHODS}
\noindent{\bf \textit{C. elegans} strain preparation.}
The Wild-type N2 Bristol strain, the \textit{mec-4(e1611)} mutant strain\cite{Chalfie:1981vd} and ZX899 strain (\textit{lite-1(ce314); ljIs123[mec-4p::ChR2, unc-122p::RFP]})\cite{Stirman:2011fb}, in which the extrachromosomal array \textit{zxEx621} from the original ZX899 strain were removed, were used in this study. For single-worm tracking, the AVA neuron of \textit{C. elegans} were marked by expressing fluorescent protein TagRFP under the control of the \textit{flp-18} promoter~\cite{Rogers:2003ju} according to the standard protocol of germline transformation\cite{Mello:1991tr}. These strains were maintained and handled by standard methods\cite{Brenner:1974wn}. 

\noindent{\bf Cultivation with DFA medium in glass bottle.} Large numbers of \textit{C.~elegans} dauer larvae were obtained by cultivating worms in DFA medium~\cite{Tanaka:2010hz,united1981monoxenic}. A glass sample bottle (Toseiyoki) containing 2 g of powdered dog food (VITA-ONE, Nihon Pet Food) and 5 ml of 1~\% agar medium was autoclaved and cooled to room temperature. Then, several drops of \textit{E. coli} OP50 suspension in Luria-Bertani broth were added to the surface of the medium. To prepare the worms for inoculation into the DFA medium, four well-fed wild-type adult worms were initially deposited on a 60 mm plate (Thermo Scientific) containing 14 ml of NGM with agar, on which \textit{E. coli} OP50 was seeded~\cite{Brenner:1974wn}. F1 worms were grown to starvation in the NGM plate at 23ºC for 4 days. The starved worms were collected from four NGM plates using autoclaved water and then inoculated into DFA medium in a glass bottle, which was then incubated at 23ºC for 7 to 10 days. The temperature of the bottom surface was changed from 23ºC to 25ºC using a Peltier temperature controller unit (Ampere). The network structures on the inner surfaces of the two glass bottles were recorded with USB-controlled CCD cameras (Sentech), which were each coupled to a 25 mm focal-length and C-mount machine vision lens (Azure) and a C-mount adapter (thickness of 5 mm). The frame rate was less than 0.20 frames \si{\per\second}. 

\noindent{\bf Cultivation with DFA on NGM plate and attraction strength control.} 
Before inoculation with worms, small amounts (approximately 0.5 g) of DFA medium were autoclaved in a glass beaker and transferred onto the centre of an NGM plate on which \textit{E. coli} OP50 was seeded. Then, starved worms collected from two NGM plates were inoculated onto the DFA on the NGM plate. Worms propagated at 23ºC climbed up to the lid of the plate for approximately 10 days.

On the day of the experiment, a Petri plate was placed onto an aluminium plate on the stage of an Olympus SZX7 microscope, which was kept at 23ºC by a Peltier temperature controller unit (Vics).  The Petri plate was maintained under this condition for 5 min before the image acquisition. Then, the temperature of the bottom of the Petri plate was increased from 23ºC to 26ºC to change the humidity inside the plate. Image acquisition of the inner surface of the plate lid was performed by the DP74 colour camera (Olympus) at 1.0 frame \si{\per\second}. A Plan Achromat objective lens (x 0.5, NA = 0.05; Olympus) was used as a low-magnification objective. The acquired images were saved in the Tagged Image File format.

\noindent{\bf Reconstruction of \textit{C. elegans} network on agar surface.} Worms were propagated with DFA on NGM in a Petri plate. The worms that moved to the lid of the plate were collected with autoclaved water and washed once. The concentration of purified worms in water was estimated by counting the number of worms in a part of the worm suspension. Based on this estimation, several worm suspensions were prepared with varying concentrations in the range of 500 worms \si{\per\micro\litre} (100,000 worms in 200 \si{\micro\litre} of water) to 2,000 worms \si{\per\micro\litre} (400,000 worms in 200 \si{\micro\litre} of water). The worm suspension was dropped onto a surface of food-free NGM in a Petri plate. Images of the agar surfaces were captured by a WSE-6100H LuminoGraph image analyser (ATTO). 

\noindent{\bf Single worm tracking in a bundle.} The position of a worm in a bundle was often impossible to measure because the worms in the bundle were too crowded. Therefore, a few head neurons of worms were marked by expressing the fluorescent protein TagRFP (Fig. 3b). The sparse TagRFP expression made it easy to locate individual worms. Images of worms on a plastic surface were taken by a Leica MZ10F fluorescence microscope equipped with a Planapochromatic objective lens (x 1.0, NA = 0.125; Leica Microsystems) at 15 frames \si{\per\second}. The movement directions of 39 worms in a bundle for 10 s were manually identified based on the positions of the fluorescence. The pie chart in Fig. 3c was created by analysing all movement directions with Mathematica 9.0 (Wolfram).

\noindent{\bf Optogenetics.} Optogenetic experiments were performed with a blue light-activated cation channel, channelrhodopsin-2 (ChR2), which has been used to noninvasively control the activity of well-defined neuronal populations~\cite{Boyden:2005cd,Deisseroth:2015dw}. ZX899 is the light-insensitive mutant \textit{lite-1(ce314)} that carries ChR2 under the control of the \textit{mec-4} promoter~\cite{Ward:2008kw,Stirman:2011fb}. ChR2 is expressed in the six touch neurons, and the activation of these neurons by blue light illumination has been known to competitively drive reversal and accelerate forward movements as major and minor responses, respectively~\cite{Stirman:2011fb}. Blue light illumination not only accelerated the movement of worms (approximately 78.9\%) that were already moving before illumination but also stimulated all inactive halted worms (approximately 21.1\%) to move actively. All worms harboured ChR2 because the extrachromosomal array including the \textit{mec-4p::ChR2} DNA was integrated into the chromosomes~\cite{Stirman:2011fb}. Worms were cultivated at 23ºC with DFA on NGM in a Petri plate under dark conditions. 40 \si{\micro\liter} of 50 \si{\micro\Molar} ATR (Sigma-Aldrich), the cofactor of ChR2, was poured onto the DFA before cultivation.

Worms climbing onto the lid of a Petri plate were illuminated on the stage of an Olympus SZX7 microscope, which was maintained at 23ºC. The Petri plate was maintained under these conditions for 5 min before blue light illumination. For ChR2 activation, a 100 W mercury lamp (Olympus) was used to deliver blue light, filtered with an SZX-MGFP set (Olympus). The illumination time was precisely controlled using a BSH2-RIX-2-1 electromagnetic shutter system (SIGMAKOKI). Image acquisition of the inner surface of the plate lid was performed by a DP74 colour camera (Olympus).
 
\noindent{\bf Estimation of parameters} To estimate the mean velocity ($v_{0}$), the mean rotation rate ($\omega_{0}$), the standard deviation of roatation rate ($\sigma_{\omega}$), and the correlation time of rotation rate ($\tau$), the behaviours of sparsely isolated worms on the inner surface of the lid of a Petri plate were recorded using Olympus cellSens software and an SZX7 microscope equipped with the DP74 camera. From the obtained images, the centroid of 38 \textit{C.~elegans} at 60 consecutive time points with an interval $\Delta t = 0.21669$ s were measured. $v_{0}$ was estimated from the average of the distance between two consecutive positions. Let the position of the $i$-th \textit{C.~elegans} at $t$ be $(x^i(t), y^i(t))$. The direction of motion, $\theta^i(t)$, was calculated as the direction of $(x^i(t + \Delta t) - x^i(t), y^i(t + \Delta t) - y^i(t))$. Then, $\omega_0$ was estimated as $\left<\theta^i(t_e) - \theta^i(0)\right> / t_e$, where $t_e = 59 \times \Delta t$, and $\left<\cdot\right>$ means the average over $i$. Here, $\tau$  is the relaxation time of the change rate of $\theta^i(t)$ to $\omega_0$. If $\theta^{i}(t)=\int {\rm d} t ( \omega_{0}+(\omega^{i}_{t=0}-\omega_{0})e^{-t/\tau})=\theta_{0}^{i}+\omega_{0}t+(\omega^{i}_{t=0}-\omega_{0})\tau \left( 1-e^{-t/\tau}\right)$,
\begin{align*}
&\left< \frac{\theta^{i}(t_{\rm e})-\theta^{i}(0)-\omega_{0}t_{\rm e}}{|\theta^{i}(t_{\rm e})-\theta^{i}(0)-\omega_{0}t_{\rm e}|}(\theta^{i}(t)-\omega_{0}t) \right>\\
=&\left<\frac{\theta^{i}(t_{\rm e})-\theta^{i}(0)-\omega_{0}t_{\rm e}}{|\theta^{i}(t_{\rm e})-\theta^{i}(0)-\omega_{0}t_{\rm e}|}\theta^{i}_{0}\right>+\left<|\omega^{i}_{t=0}-\omega_{0}|\right>\tau\left(1- e^{-\frac{t}{\tau}}\right)\\
=&\left<\frac{\theta^{i}(t_{\rm e})-\theta^{i}(0)-\omega_{0}t_{\rm e}}{|\theta^{i}(t_{\rm e})-\theta^{i}(0)-\omega_{0}t_{\rm e}|}\theta^{i}_{0}\right>+\sqrt{\frac{2}{\pi}}\sqrt{\left<\left(\omega_{t=0}^{i}-\omega_{0}\right)^{2}\right>}\tau\left(1- e^{-\frac{t}{\tau}}\right)
\end{align*}
on the assumption that $\omega_{t=0}^{i}-\omega_{0}$ were normally distributed and $(\theta^{i}(t_{\rm e})-\theta^{i}(0)-\omega_{0}t_{\rm e})(\omega^{i}_{t=0}-\omega_{0})>0$. $\tau$ and $\sigma_{\omega}$ were estimated from the fitting of $\left<\frac{\theta^{i}(t_{\rm e})-\theta^{i}(0)-\omega_{0}t_{\rm e}}{|\theta^{i}(t_{\rm e})-\theta^{i}(0)-\omega_{0}t_{\rm e}|}(\theta^{i}(t)-\omega_{0}t) \right>$ to $\theta_{0} + \sqrt{2/\pi}\sigma_{\omega} \tau (1- e^{-\frac{t}{\tau}})\approx \theta_{0} + \sqrt{2/\pi}\sigma_{\omega} \left(t -\frac{t^{2}}{2\tau}\right)$.

\noindent {\bf Measurement of the size of areas surrounded by bundles.} To measure the area sizes, the moving averaged images of collective motion were first binarized using Fiji (https://fiji.sc). The window width was 50 s in the \textit{C. elegans} case and 100 time units in the model case. The binarized images were skeletonized using Matlab. The sizes of the areas surrounded by bundles were measured by counting the pixels surrounded by lines in the skeletonized images.

\bibliographystyle{naturemag}


\noindent{\bf Supplementary Information} is available in the online version of the paper.

\clearpage    

\noindent{\bf Acknowledgements}
We thank the \textit{Caenorhabditis} Genetics Center for sharing strains and the member of T. Ishihara (Kyushu University) and T. Nakagaki (Hokkaido University) laboratories for valuable comments on this manuscript and discussion. T.~S., H.~I. and K.~H.~N. were supported by the Japan Society for the Promotion of Science (grant no. 17KT0016). T.~S. was supported by the Japan Science and Technology Agency under Precursory Research for Embryonic Science and Technology (PRESTO) and by the Mochida Memorial Foundation for Medical and Pharmaceutical Research.

\noindent{\bf Author Contributions}
H.~I. discovered the phenomenon of dynamical network formation by \textit{C. elegans} with the help of T.~S. T.~S., H.~I., and K.~H.~N. conceived the experiments, and T.~S. performed the experiments with M.N. K.~H.~N. conceived and analysed the simulations. T.~S., H.~I., and K.~H.~N. analysed the obtained images and wrote the paper. 

\noindent{\bf Author Information} Reprints and permissions information is available at \url{www.nature.com/reprints}. The authors declare no competing financial interests. Readers are welcome to comment on the online version of the paper. Correspondence and requests for materials should be addressed to T.~S.~(tsugi@belle.shiga-med.ac.jp), H.~I.~(hito@design.kyushu-u.ac.jp), K.~H.~N.~(k-nagai@jasit.ac.jp).
 
\section*{Figure}

\begin{figure}
    \centering
    \includegraphics[width=\linewidth]{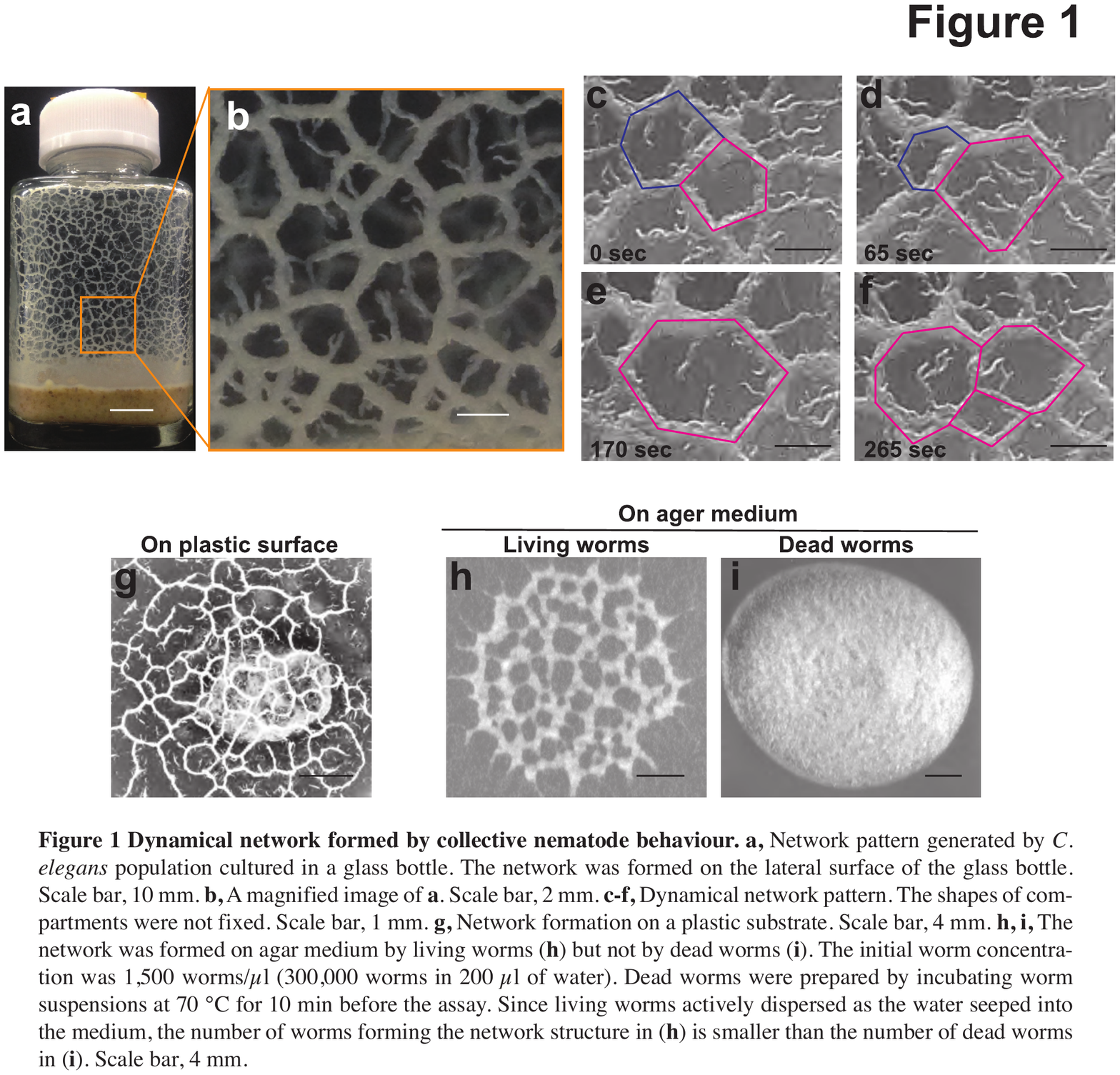}
\end{figure}
\begin{figure}
    \centering
    \includegraphics[width=\linewidth]{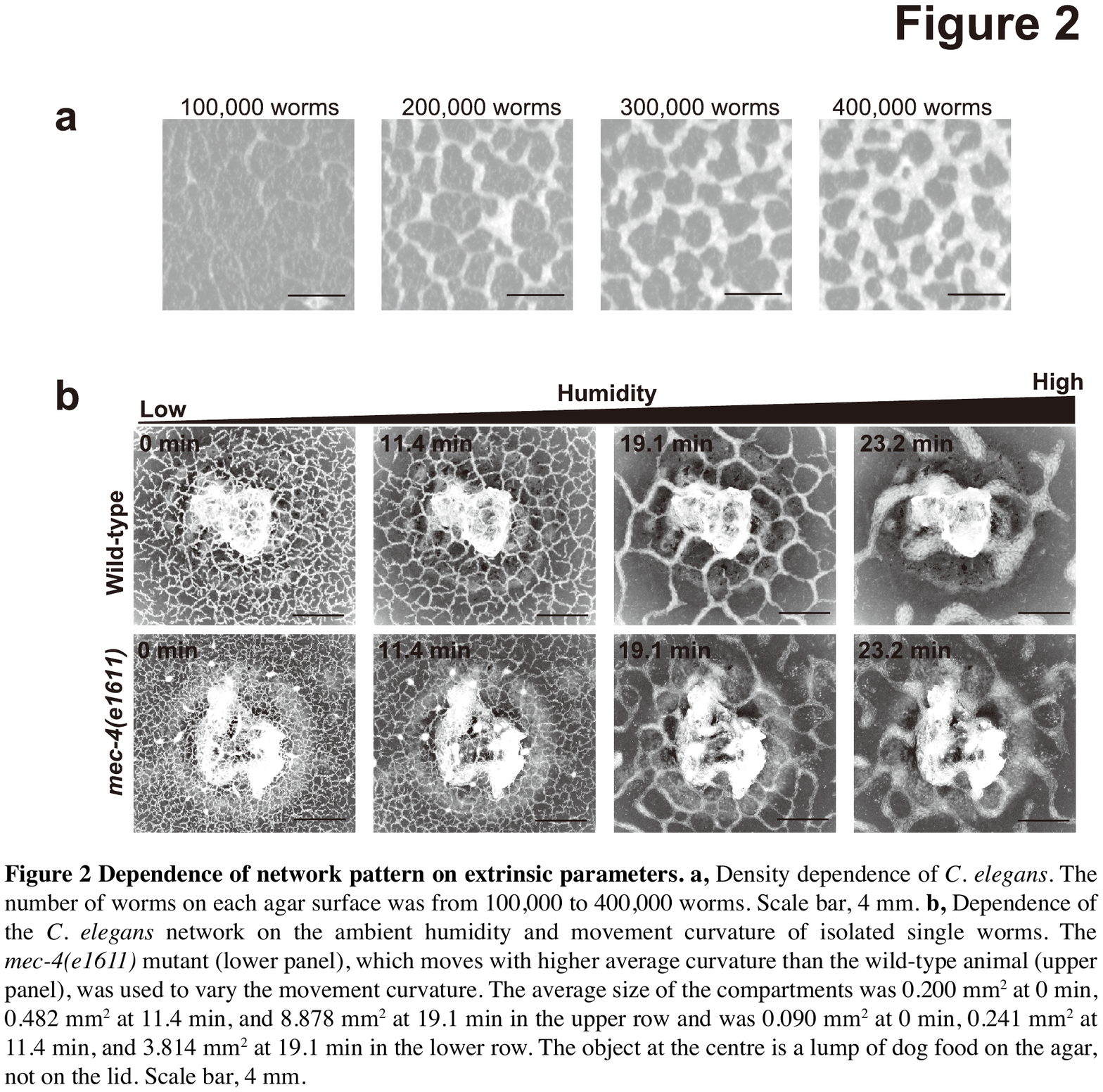}
\end{figure}
\begin{figure}
    \centering
    \includegraphics[width=\linewidth]{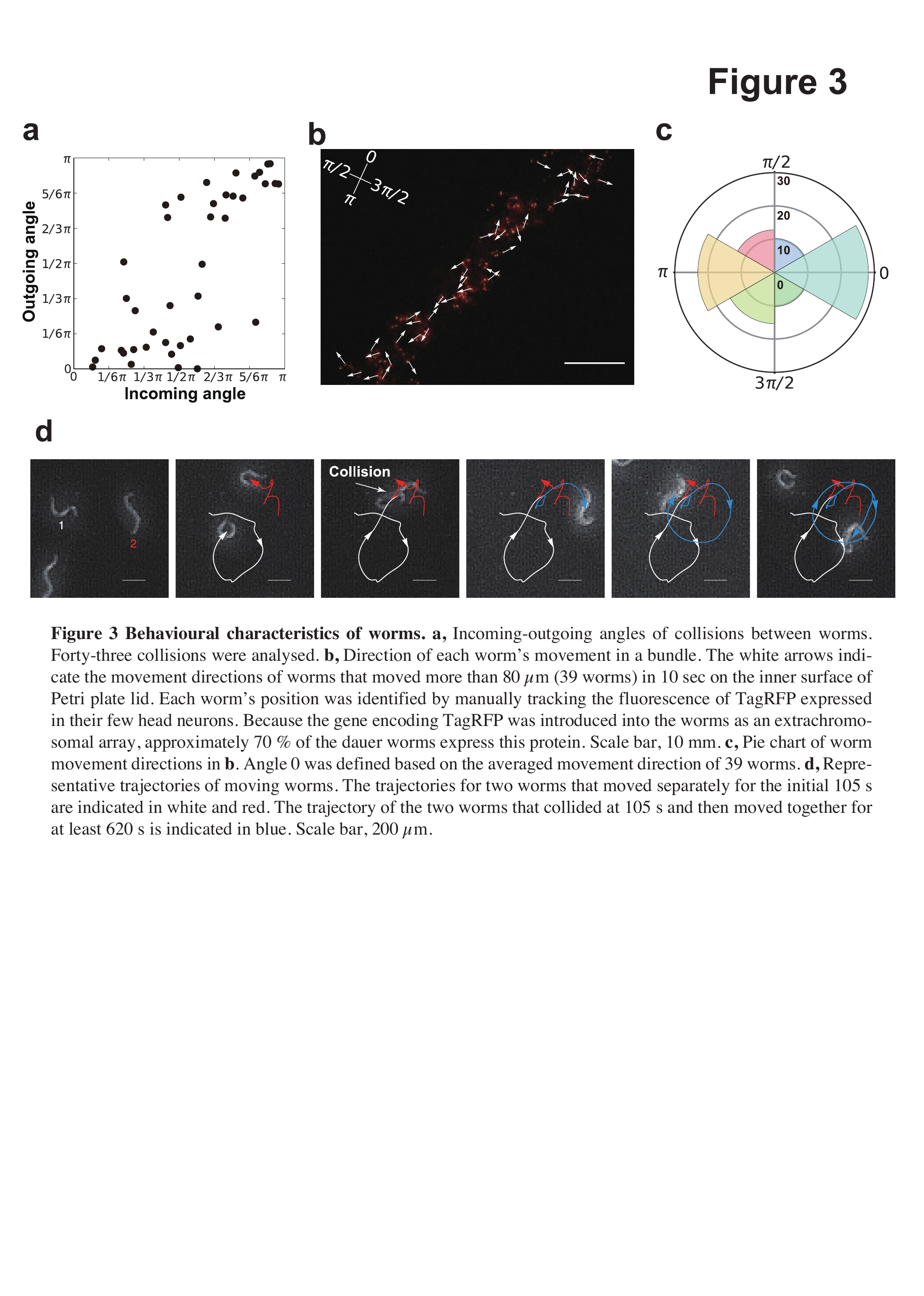}
\end{figure}
\begin{figure}
    \centering
    \includegraphics[width=\linewidth]{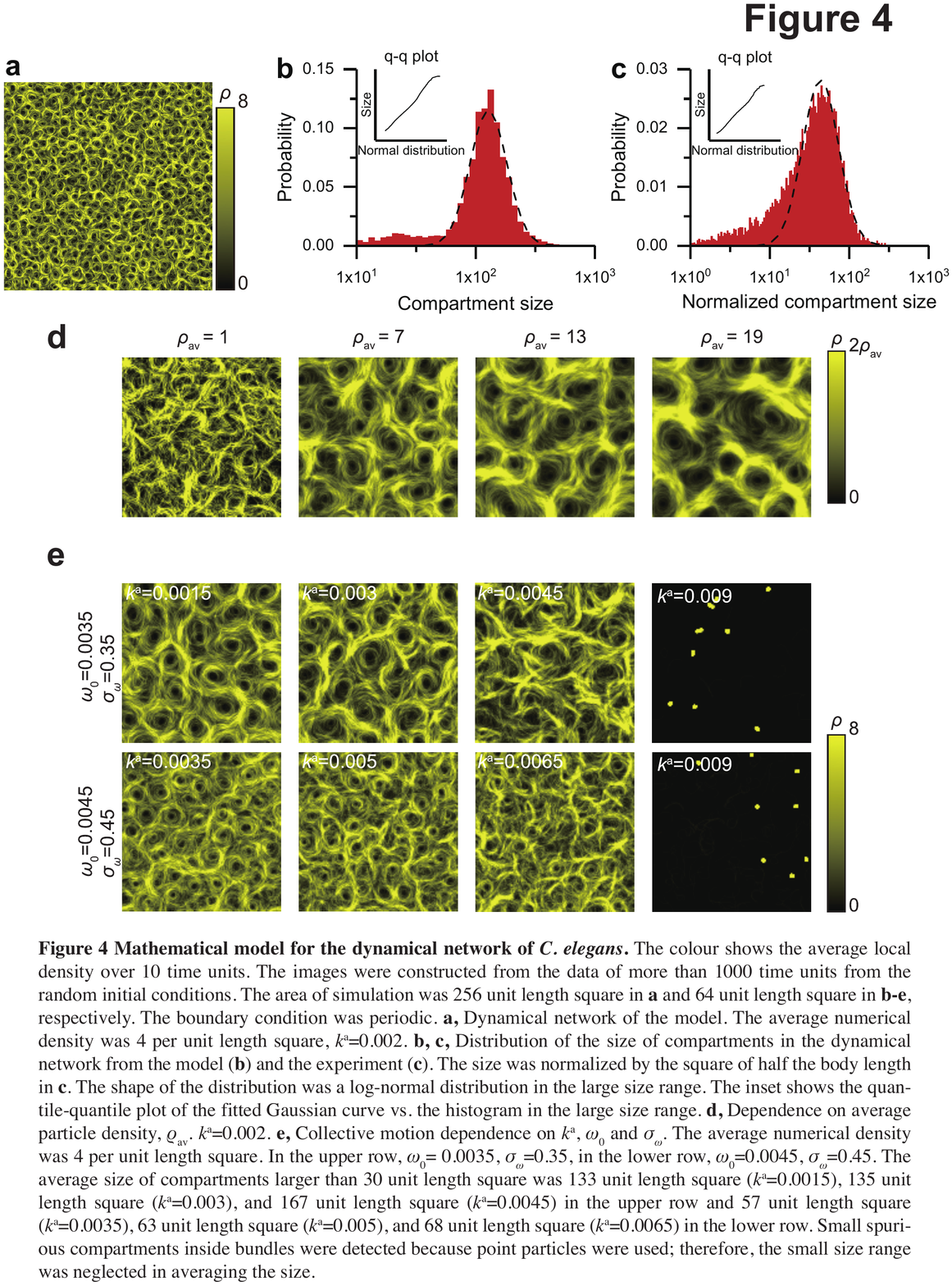}
\end{figure}
\begin{figure}
    \centering
    \includegraphics[width=\linewidth]{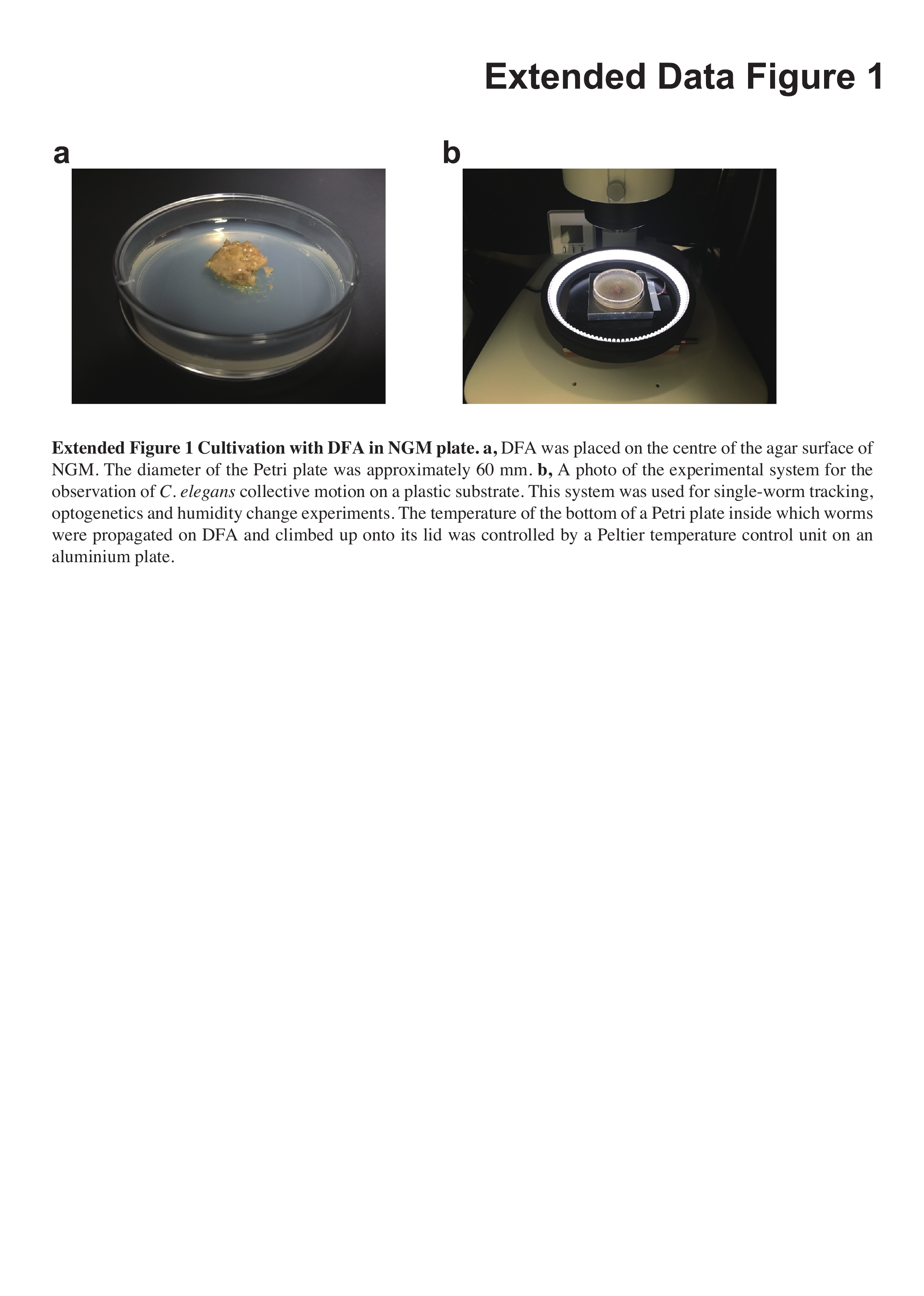}
\end{figure}

\end{document}